\documentclass[prd,showpacs]{revtex4}
\usepackage{amssymb}
\usepackage{amsmath}
\usepackage{amsfonts}
\usepackage{array}
\usepackage{dcolumn}
\usepackage{longtable}
\usepackage{hyperref}

\begin{document}

\title{Nonlinear coupled Alfv\'{e}n and gravitational waves}
\author{Andreas K\"{a}llberg}
\email[E-mail: ]{andreas.kallberg@physics.umu.se}
\author{Gert Brodin}
\email[E-mail: ]{gert.brodin@physics.umu.se}
\author{Michael Bradley}
\email[E-mail: ]{michael.bradley@physics.umu.se}
\affiliation{Department of Physics, Ume\aa\ University, SE-901 87 Ume\aa, Sweden}
\date{\today}

\begin{abstract}
In this paper we consider nonlinear interaction between gravitational and
electromagnetic waves in a strongly magnetized plasma. More specifically, we
investigate the propagation of gravitational waves with the direction of
propagation perpendicular to a background magnetic field, and the coupling
to compressional Alfv\'{e}n waves. The gravitational waves are
considered in the high frequency limit and the plasma is modelled by a
multifluid description. We make a self-consistent, weakly nonlinear analysis of the
Einstein-Maxwell system and derive a wave equation for the coupled
gravitational and electromagnetic wave modes. A WKB-approximation is then
applied and as a result we obtain the nonlinear
Schr\"{o}dinger equation for the slowly varying wave amplitudes. The
analysis is extended to 3D wave pulses, and we discuss the applications to
radiation generated from pulsar binary mergers. It turns out that the
electromagnetic radiation from a binary merger should experience a focusing
effect, that in principle could be detected.
\end{abstract}


\maketitle

\section{Introduction}

Recently much work has been devoted to the study of gravitational waves,
largely due to the increased possibility of detection by facilities in
operation such as LIGO (Laser Interferometer Gravitational Wave Observatory), or by ambitious detector projects under development such as
LISA (Laser Interferometer Space Antenna) \cite{maggiore}. The possibility of
interaction between electromagnetic and gravitational fields has also led
to alternative proposals for gravitational wave detectors, see e.g. Refs. 
\cite{BrodinMarklund2003}-\cite{Picasso2003} and references therein. Closer
to the source, in an astrophysical context, the gravitational waves often
propagate in a plasma medium, and the amplitudes can be much larger, which
increases the number of possible interaction mechanisms, see e.g. Refs. \cite
{macedo}-\cite{Anile99}. Linear gravitational wave theory in a magnetized
plasma has been studied by for example Refs \cite{macedo}-\cite{Servin2001}%
, including the backreaction from the plasma on the gravitational wave. In
Refs. \cite{brodinmarklund}-\cite{bmd2} the authors have studied nonlinear
responses to the gravitational wave by the plasma medium, although the
backreaction has been neglected. The nonlinear response gives raise to
effects such as parametric instabilities \cite
{Papadouplous2001,bmd2,Balakin2003,Servin2000}, large density fluctuations 
\cite{bms,ignatev} and photon acceleration \cite{bms}. The application of
gravitational wave processes to astrophysics has been discussed by for
example Refs. \cite{bmd1}-\cite{Mosquera2002}, and to cosmology by Refs. 
\cite{Papadoupolus2002}-\cite{Hogan2002}. A number of works studying
nonlinear propagation of gravitational waves including the backreaction from
the plasma have also been written, see e.g. \cite
{ignatev,Balakin2003,servinbrodin}.

In Refs. \cite{Servin2003,Mendonca2003} \ geometrical nonlinearities from
the Einstein tensor were considered, and a nonlinear evolution equation was
derived. However, it was found that the nonlinear coefficient was
proportional to the small difference of the phase velocity and the velocity
of light in vacuum. In the present paper we neglect nonlinearities from the
Einstein tensor and instead focus on the nonlinear response from matter. In
particular we consider coupled gravitational and electromagnetic waves
propagating in a magnetized plasma, with the direction of propagation
perpendicular to the background magnetic field. For this wave coupling to be
efficient, the interaction should be (almost) resonant, i.e. the
electromagnetic wave propagation velocity in the plasma should be close to
the speed of light. This increases the interaction strength and allows us to
neglect the effect of the background curvature, in comparison with the direct interaction with matter.

The plasma is modelled by a multifluid description, and we perform a
self-consistent weakly nonlinear analysis of the Einstein-Maxwell system of
equations for 1D spatial variations in the high-frequency limit. This system
is reduced to a single nonlinear wave equation for the coupled gravitational
and electromagnetic waves. We then apply a WKB-approximation to this wave
equation, and it turns out that the slowly varying wave amplitude obey the
well-known nonlinear Schr\"{o}dinger (NLS) equation \cite{doddeilbeck}. In section \ref{3dsect}, the analysis is expanded to include a 3D spatial
dependence, allowing us to consider diffraction and/or nonlinear
self-focusing of the wave. For certain conditions the sign of the
nonlinear coefficient is of focusing type, which for sufficient initial
amplitudes implies solutions that undergo wave collapse. The conditions for
collapse and the possible applications to astrophysics are discussed.

\section{Overview of the approximation scheme}

The problem treated in this paper contains a number of small parameters used in the different approximations and expansions made throughout the paper. The aim of this section is to give an overview of these parameters, at what stages of the calculations they are introduced, and to what order the expansions need to be made. 

In section \ref{basic}, we introduce the small gravitational wave amplitudes, $h_{+,\times}$, and in calculating Maxwell and fluid equations, (\ref{me2})-(\ref{fleq3}), we only keep terms to $\mathcal{O}(h_{+,\times})$. However, for the case when the gravitational wave interacts with an almost resonant Alfv\'en wave, nonlinearities become important. The reason is that the resonance magnifies $| \delta {\bf B}|/|{\bf B}_{0}|$ where $\delta{\bf B}$ and ${\bf B}_{0}$ are the wave magnetic field and background magnetic field, respectively. The magnification is of the order $| \delta {\bf B}|/|{\bf B}_{0}|\sim h_{+}\omega/\delta\omega$ where $\delta\omega$ is the frequency mismatch as compared to the Alfv\'en mode. Therefore, we keep terms up to order $\mathcal{O}(| \delta {\bf B}|^3/|{\bf B}_{0}|^3)$, in the expansion of the wave equation (\ref{waveqfin}), in order to obtain the nonlinear amplitude modulation. Other nonlinearities originating from the Einstein tensor and the effective currents in (\ref{me1})-(\ref{momeq1}) will not be magnified due to the resonance, and may be neglected in our treatment, see also the discussion following Eq. (\ref{wgwmetric}).

Furthermore, in section \ref{basic} we introduce the high frequency approximation, valid for wavenumbers $k\gg 1/r_{c}$, where $r_{c}$ is a characteristic radius of curvature of the background. This condition may also be written as the small parameter $\kappa B_{0}^{2}/k^{2}\ll 1$, where $\kappa =8\pi G$ (we use
units where $c=\mu_{0}=\epsilon_{0}=1$). We have to keep terms to $\mathcal{O}(\kappa B_{0}^{2}/k^{2})$, in order to include the weak coupling between the gravitational mode and the Alfv\'en mode.

In section \ref{propsect} we first use the approximations $\partial_{t}\approx-\partial_{z}$, except in the linear wave operators, in order to simplify our system of equations. This is justified in the high frequency approximation. Furthermore, we use $\partial_{t}\ll\omega_{c}$, i.e. we assume that the frequency of variations of fields is much smaller than the cyclotron frequency of all particle species in the plasma. In the expansion of Eq. (\ref{v3eq}), we must include terms to $\mathcal{O}(\partial_{t}/\omega_{c})$. This is because the equations of motion to zeroth order in $\partial_{t}/\omega_{c}$ gives just the ${\bf E}\times{\bf B}$ drift of all particle species, and we must include terms of one order higher to obtain a nonzero current in the plasma.

Next, in section \ref{nlssect}, we introduce a weakly modulated wave (WKB-approximation), with the underlying assumption that the characteristic length scale for amplitude modulations obeys $L_{\parallel}\gg\lambda$. In deriving Eq. (\ref{orgeq}), we need terms to $\mathcal{O}(\lambda^{2}/L_{\parallel}^{2})$ in order to include the dispersive term in (\ref{orgeq}). We also use $1/C_A^2\ll 1$, where $C_A$ is the Alfv\'en velocity, and only keep terms to lowest non-vanishing order to simplify the equations. Then, in section \ref{3dsect}, we extend the problem to include small variations in the directions perpendicular to the direction of propagation. We apply the condition $L_{\perp}\gg\lambda$, where $L_{\perp}$ is the characteristic length for perpendicular variations. In subsequent calculations, we must keep terms to $\mathcal{O}(\lambda^{2}/L_{\perp}^{2})$ to obtain the diffractive term in Eq. (\ref{NLS-3D}), see the discussion at the beginning of section \ref{3dsect} for more details.

The parameters in this section are in principle independent of each other, with the sole exception that $\delta\omega/\omega\sim1/C_{A}^{2}$, which is our main motivation for considering the regime $C_{A}^{2}\gg 1$. In addition to these parameters there are a few other small parameters introduced throughout the paper, but those quantities are related in different ways to the ones presented in this section, and the relations are explained in the text. The only restrictions on the parameters in this section, besides their smallness, is due to physical applicability, i.e. the values of the parameters should be chosen to fit environments close to emitters of reasonably strong gravitational waves. An example of such parameter values is presented in the final section.

%vided the Alfv\'en velocity obeys $C_{A}^{2}\gg 1$, see Eq. (\ref{waveq}).In the WKB-approximation we include terms to lowest non-vanishing order in the parameter %$1/C_{A}^{2}$, %in order to obtain corrections to the dispersion relations for the uncoupled waves.

\section{Basic equations}

\label{basic}

The gravitational and electromagnetic fields are governed, respectively, by
the Einstein field equations 
\begin{equation}
G_{ab}=\kappa T_{ab}  \label{efe}
\end{equation}
\noindent and the Maxwell field equations 
\begin{eqnarray}
\nabla _{a}F^{ab} &=&j^{b}  \label{mfe1} \\
\nabla _{a}F_{bc}+\nabla _{b}F_{ca}+\nabla _{c}F_{ab} &=&0  \label{mfe2}
\end{eqnarray}
\noindent where $G_{ab}$ is the Einstein tensor, $T_{ab}$ is the energy-momentum tensor, $F_{ab}$ is the
electromagnetic field tensor, $j^{b}$ is the 4-current density and $\nabla $
denotes covariant differentiation. We use the spacelike signature $(-+++)$
for the metric.

For the plasma we use a multifluid description, where the plasma is
seen as a number of charged fluids, one for each plasma species, and
collisions between particles are neglected. The total energy-momentum tensor
is then $T_{ab}=T_{ab}^{(fl)}+T_{ab}^{(em)}$. Here the fluid part is 
\begin{equation}
T_{ab}^{(fl)}=\sum_{s}\left[(\mu _{(s)}+p_{(s)})u_{(s)a}u_{(s)b}+p_{(s)}g_{ab}\right]
\end{equation}
where $\mu_{(s)}, p_{(s)}$ and $u_{(s)a}$ are the internal energy density, pressure and four-velocity for each plasma species, $s$, $g_{ab}$ is the metric tensor and the electromagnetic
part is 
\begin{equation}
T_{ab}^{(em)}=F_{a}^{\ c}F_{bc}-{\textstyle\frac{1}{4}}g_{ab}F^{cd}F_{cd}
\end{equation}
In the absence of collisions, the evolution equations for each fluid species
can be written \cite{fotnot1} 
\begin{equation}
\nabla _{b}T_{(s)}^{ab}=F^{ab}j_{b(s)}  \label{conseq1}
\end{equation}
as is consistent with Eq. (\ref{efe})

With these preliminaries, we now follow the approach applied in Refs. \cite
{bms,bmd2,bmd1,servinbrodin,elliselst,bmd3} and introduce an observer
four-velocity, $V^{a}$, so that the electromagnetic field can be decomposed
relative to this into an electric and a magnetic part, $E_{a}=F_{ab}V^{b}$ and $%
B_{a}={\textstyle\frac{1}{2}}\epsilon _{abc}F^{bc}$ respectively. Here $%
\epsilon _{abc}=V^{d}\epsilon _{abcd}$ where $\epsilon _{abcd}$ is the
4-dimensional volume element with $\epsilon _{0123}=\sqrt{|\mathrm{det}\
g_{ab}|}$. Next we introduce an orthonormal frame (ONF) with basis $\left\{ 
\mathbf{e}_{a}=e_{a}^{\mu}\partial_{x^{\mu}}\right\}$, where $\mathbf{e}_{0}=%
\mathbf{V}=V^{a}\mathbf{e}_{a}$, (i.e. $V^{a}=\delta^{a}_{0}$), and write
the fluid 4-velocity as $u^{a}=(\gamma ,\gamma \mathbf{v})$, where $\gamma
=(1-v_{\alpha }v^{\alpha })^{-\frac{1}{2}}$, $\alpha=1,2,3$, and $\mathbf{v}$
is the fluid three velocity. Dividing$\ $\ the four-current$\ j^{a}=%
\displaystyle\sum_{s}q_{(s)}n_{(s)}u_{(s)}^{a}$ in the same manner, the
Maxwell equations (\ref{mfe1}, \ref{mfe2}) and fluid equations (\ref{conseq1}%
) can be written \cite{bmd1,servinbrodin,bmd3} 
\begin{eqnarray}
\nabla \cdot \mathbf{E} &=&\rho +\rho _{E}  \label{me1} \\
\nabla \cdot \mathbf{B} &=&\rho _{B} \\
\mathbf{e}_{0}\mathbf{E}-\nabla \times \mathbf{B} &=&-\mathbf{j}-\mathbf{j}%
_{E} \\
\mathbf{e}_{0}\mathbf{B}+\nabla \times \mathbf{E} &=&-\mathbf{j}_{B} \\
\mathbf{e}_{0}(\gamma n)+\nabla \cdot (\gamma n\mathbf{v}) &=&\Delta n
\label{fleq1} \\
(\mu +p)(\mathbf{e}_{0}+\mathbf{v}\cdot \nabla )\gamma \mathbf{v} &=&-\gamma
^{-1}\nabla p-\gamma \mathbf{v}(\mathbf{e}_{0}+\mathbf{v}\cdot \nabla )p+qn(%
\mathbf{E}+\mathbf{v}\times \mathbf{B})+(\mu +p)\mathbf{g}  \label{momeq1}
\end{eqnarray}
where we have omitted the fluid species index, $s$, and it is understood
that we have one pair of fluid equations (\ref{fleq1}, \ref{momeq1}) for
each plasma fluid species. We have introduced the 3-vector notation $\mathbf{%
E}\equiv (E^{\alpha })=(E^{1},E^{2},E^{3})$ etc. and $\nabla \equiv (\mathbf{%
e}_{1},\mathbf{e}_{2},\mathbf{e}_{3})$. The charge density is $\rho =%
\sum_{s}q_{(s)}\gamma n_{(s)}$ and $n$ is the proper particle
number density. The effective charges, currents and forces, originating from
the inclusion of the gravitational field, are given by 
\begin{eqnarray}
\rho _{E} &\equiv &-\Gamma _{\beta \alpha }^{\alpha }E^{\beta }-\epsilon
^{\alpha \beta \gamma }\Gamma _{\alpha \beta }^{0}B_{\gamma }  \label{roe1}
\\
\rho _{B} &\equiv &-\Gamma _{\beta \alpha }^{\alpha }B^{\beta }+\epsilon
^{\alpha \beta \gamma }\Gamma _{\alpha \beta }^{0}E_{\gamma } \\
\mathbf{j}_{E} &\equiv &\left[ -(\Gamma _{0\beta }^{\alpha }-\Gamma _{\beta
0}^{\alpha })E^{\beta }+\Gamma _{0\beta }^{\beta }E^{\alpha }-\epsilon
^{\alpha \beta \gamma }(\Gamma _{\beta 0}^{0}B_{\gamma }+\Gamma _{\beta
\gamma }^{\delta }B_{\delta })\right] \mathbf{e}_{\alpha } \\
\mathbf{j}_{B} &\equiv &\left[ -(\Gamma _{0\beta }^{\alpha }-\Gamma _{\beta
0}^{\alpha })B^{\beta }+\Gamma _{0\beta }^{\beta }B^{\alpha }+\epsilon
^{\alpha \beta \gamma }(\Gamma _{\beta 0}^{0}E_{\gamma }+\Gamma _{\beta
\gamma }^{\delta }E_{\delta })\right] \mathbf{e}_{\alpha } \\
\Delta n &\equiv &-\gamma n(\Gamma _{0\alpha }^{\alpha }+\Gamma
_{00}^{\alpha }v_{\alpha }+\Gamma _{\beta \alpha }^{\alpha }v^{\beta }) \\
\mathbf{g} &\equiv &-\gamma \left[ \Gamma _{00}^{\alpha }+(\Gamma _{0\beta
}^{\alpha }+\Gamma _{\beta 0}^{\alpha })v^{\beta }+\Gamma _{\beta \gamma
}^{\alpha }v^{\beta }v^{\gamma }\right] \mathbf{e}_{\alpha }  \label{gforce1}
\end{eqnarray}
where $\Gamma _{bc}^{a}$ are the Ricci rotation coefficients associated with
the tetrad $\left\{ \mathbf{e}_{a}\right\} $.

From now on we will assume that the plasma is cold, so that we can neglect
the pressure terms in Eq. (\ref{momeq1}) and let $\mu_{(s)} =m_{(s)}n_{(s)}$, where $m_{(s)}$ is the mass of each particle species. We will also
apply the high frequency approximation \cite{isaacson} for the gravitational
waves. Thus we will assume that the background gravitational
field and the unperturbed plasma and electromagnetic fields fulfill
Einstein's field equations (\ref{efe}), and we introduce perturbations to
the metric, $g_{\mu\nu}=g_{\mu\nu}^{(0)}+h_{\mu\nu}$, (greek indices are coordinate indices), and to the energy-momentum tensor, $%
T_{ab}=T_{ab}^{(0)}+\delta T_{ab}$, which should then fulfill 
\begin{equation}
\delta G_{ab}=\kappa \delta T_{ab}  \label{pertefe}
\end{equation}
where $\delta G_{ab}$ is the (linearized) perturbation to the Einstein
tensor caused by the metric perturbation.

It was recently shown by Ref. \cite{servinbrodin} that in the high frequency
approximation the gravitational wave can be taken to be in the transverse
and traceless (TT) gauge even in the presence of matter. In this gauge the
metric of a linearized gravitational wave propagating in the z-direction is
given by \cite{landlif} 
\begin{equation}
ds^{2}=-dt^{2}+(1+h_{+})dx^{2}+(1-h_{+})dy^{2}+2h_{\times }dxdy+dz^{2}
\label{wgwmetric}
\end{equation}
where $h_{+}\equiv h_{+}(z,t)$ and $h_{\times }\equiv h_{\times }(z,t)$
denotes the two polarization modes of weak gravitational waves in the
TT-gauge, and $\lvert h_{+}\rvert ,\lvert h_{\times }\rvert \ll 1$. The
quantities $h_{+}$ and $h_{\times }$ differ from the corresponding
quantities in the vacuum case, $h_{+}\equiv h_{+}(z-t),h_{\times }\equiv
h_{\times }(z-t)$, because of the weak interaction with the plasma. However,
this difference is very small, which will allow us to use $\partial
_{t}\approx -\partial _{z}$ in many of the subsequent approximations. We note that components of the metric tensor corresponding to the
quadratically nonlinear contribution from the pseudo-energy momentum tensor,
as well as  higher order cubic nonlinearities are neglected in (\ref
{wgwmetric}).The basic motivation for this is that we neglect nonlinearities
from the Einstein tensor, and focus on the nonlinear response from matter.
When determining the regime of validity for this approach, a couple of
things should be noted

\begin{itemize}
\item  As shown by Ref. \cite{Servin2003}, for uni-directional propagation
of a weakly modulated plane wave in a flat background (in the absence of the
gravitational wave), the nonlinearities in the Einstein tensor cancel up to cubic order as
far as the evolution of the amplitude function is considered. (Although the
background curvature due to the pseudo-energy momentum tensor of the wave is
affected, this does not lead to amplitude modulations. We note that this
result is consistent with the exact PP-wave solutions where the amplitude
function is unaffected by the nonlinearities in the Einstein tensor,
although the background curvature is modified due to the wave. )  

\item  If the background in the absence of the gravitational wave is weakly
curved (with a characteristic radius of curvature $r_{c}$ fulfilling $%
kr_{c}\gg 1$, where $k$ is the gravitational wave number), nonlinearities in
the Einstein tensor lead to amplitude modulations of the order of  $\partial
_{t}h_{+,\times }\sim \omega h_{+,\times }^{3}/(k^{2}r_{c}^{2})$, see Ref.
\cite{Mendonca2003}.

\item  If there are wave perturbations in matter close to resonance (i.e.
assuming a wave mode in matter fulfilling $\omega =k+\delta \omega $, 
where $%
\delta \omega \ll \omega $), the matter response due to the gravitational
wave is magnified.  As will be shown below, for our case the nonlinear
contribution, close to resonance, to the amplitude modulation scales as $%
\partial _{t}h_{+,\times }\sim \kappa T_{0}h_{+,\times }^{3}\omega/\delta \omega ^{2}$, where $T_{0}$ is the magnitude of the background
energy-momentum tensor. We note that the nonlinear matter response dominates
over that associated with the combined effect of the background curvature
and the Einstein tensor nonlinearities, provided $k^{-2}r_{c}^{-2}\ll \kappa
T_{0} /\delta \omega ^{2}.$

\item  Using the high-frequency approximation ($kr_{c}\gg 1$, see \cite
{isaacson, grishchuk}) and assuming a weakly nonlinear response from matter,
it is always possible to use the TT-gauge \cite{servinbrodin}. As far as the
direct interaction with the medium is concerned, i.e. the contribution that is magnified due to the resonance, the background metric can
then be considered as flat (i.e. $g_{ab}^{(0)}=\eta _{ab}$), see \cite
{grishchuk}
\end{itemize} %We
%have also used the fact that, in the high frequency approximation, the
%background metric can be considered as flat (i.e. $g_{\mu\nu}^{(0)}=\eta _{\mu\nu}$%
%), provided the response from matter on the gravitational wave is almost resonant, see Refs. \cite
%{servinbrodin, isaacson, grishchuk} and references therein.

Next we choose our tetrad basis for the metric (\ref{wgwmetric}) as 
\begin{equation}
\mathbf{e}_{0}=\partial _{t}\ ,\ \mathbf{e}_{1}=(1-{\textstyle\frac{1}{2}}%
h_{+})\partial _{x}-{\textstyle\frac{1}{2}}h_{\times }\partial _{y}\ ,\ 
\mathbf{e}_{2}=(1+{\textstyle\frac{1}{2}}h_{+})\partial _{y}-{\textstyle%
\frac{1}{2}}h_{\times }\partial _{x}\ ,\ \mathbf{e}_{3}=\partial _{z}
\label{tetrad}
\end{equation}
Using (\ref{wgwmetric}) and (\ref{tetrad}) in the linearized Einstein field
equations (\ref{pertefe}), subtracting the $_{11}$ and $_{22}$, and adding
the $_{12}$ and $_{21}$ components of the equations one gets \cite{bms,
servinbrodin, grishchuk} 
\begin{eqnarray}
\left( \partial _{t}^{2}-\partial _{z}^{2}\right) h_{+} &=&\kappa \left(
\delta T_{11}-\delta T_{22}\right)  \label{ee1122} \\
\left( \partial _{t}^{2}-\partial _{z}^{2}\right) h_{\times } &=&\kappa
\left( \delta T_{12}+\delta T_{21}\right)  \label{ee1221}
\end{eqnarray}

In the basis (\ref{tetrad}) the non-zero terms in (\ref{roe1})-(\ref{gforce1}%
) are 
\begin{eqnarray}
\mathbf{j}_{E} &=&-{\textstyle\frac{1}{2}}\left( E_{1}\partial
_{t}h_{+}+B_{2}\partial _{z}h_{+}+E_{2}\partial _{t}h_{\times
}-B_{1}\partial _{z}h_{\times }\right) \mathbf{e}_{1}  \notag \\
&&+{\textstyle\frac{1}{2}}\left( E_{2}\partial _{t}h_{+}-B_{1}\partial
_{z}h_{+}-E_{1}\partial _{t}h_{\times }-B_{2}\partial _{z}h_{\times }\right) 
\mathbf{e}_{2}  \label{je2} \\
&&  \notag \\
\mathbf{j}_{B} &=&-{\textstyle\frac{1}{2}}\left( B_{1}\partial
_{t}h_{+}-E_{2}\partial _{z}h_{+}+B_{2}\partial _{t}h_{\times
}+E_{1}\partial _{z}h_{\times }\right) \mathbf{e}_{1}  \notag \\
&&+{\textstyle\frac{1}{2}}\left( B_{2}\partial _{t}h_{+}+E_{1}\partial
_{z}h_{+}-B_{1}\partial _{t}h_{\times }+E_{2}\partial _{z}h_{\times }\right) 
\mathbf{e}_{2} \\
&&  \notag \\
\mathbf{g} &=&-{\textstyle\frac{1}{2}}\gamma \left( v_{1}\partial
_{t}h_{+}+v_{1}v_{3}\partial _{z}h_{+}+v_{2}\partial _{t}h_{\times
}+v_{2}v_{3}\partial _{z}h_{\times }\right) \mathbf{e}_{1}  \notag \\
&&+{\textstyle\frac{1}{2}}\gamma \left( v_{2}\partial
_{t}h_{+}+v_{2}v_{3}\partial _{z}h_{+}-v_{1}\partial _{t}h_{\times
}-v_{1}v_{3}\partial _{z}h_{\times }\right) \mathbf{e}_{2}  \notag \\
&&+{\textstyle\frac{1}{2}}\gamma \left( (v_{1}^{2}-v_{2}^{2})\partial
_{z}h_{+}+2v_{1}v_{2}\partial _{z}h_{\times }\right) \mathbf{e}_{3}
\label{geforce2}
\end{eqnarray}

The generalization to a finite gravitational wave pulse width, i.e. including a weak spatial
dependence perpendicular to the direction of propagation, leads to
additional terms in (\ref{je2})-(\ref{geforce2}) and (\ref{ee1122})-(\ref
{ee1221}). An outline of this case is found in section \ref{3dsect}.

\section{Coupled Alfv\'{e}n and gravitational waves}

\label{propsect}

In Ref. \cite{bms} a test fluid approach was taken to show that a gravitational wave
can drive the amplitude of an electromagnetic wave to a nonlinear regime, and in Ref. \cite
{servinbrodin} a linear analysis of the interaction between a gravitational wave and the
extraordinary electromagnetic wave was made using the equations presented in section \ref
{basic}. Here we intend to take a similar approach, but guided by the
results in Ref. \cite{bms}, we will include a nonlinear response of
matter and fields, while still assuming the metric perturbation to remain
small.

We assume the presence of a background magnetic field, $\mathbf{B}_{0}=B_{0}%
\mathbf{e}_{1}$, and introduce the perturbations: $n=n_{0}+\delta n$, $%
\mathbf{B}=(B_{0}+B_{x})\mathbf{e}_{1}$, $\mathbf{E}=E_{y}\mathbf{e}%
_{2}+E_{z}\mathbf{e}_{3}$ and $\mathbf{v}=v_{y}\mathbf{e}_{2}+v_{z}\mathbf{e}%
_{3}$ \cite{fotnot2}. We also note that in the case
of gravitational waves propagating in a magnetized plasma, with the magnetic
field perpendicular to the direction of propagation, only the $h_{+}$%
-polarization part of the gravitational wave couples effectively \cite{fotnot3}
to the electromagnetic wave, see also Ref. \cite{servinbrodin}. Thus we put $h_{\times }=0$
in order to simplify the algebra. As in Ref. \cite{bms}, we assume $v_{y}\ll
1$, (because $v_{y}\sim h_{+}$), and therefore neglect terms of the type $v_{y}^{2}$ and $v_{y}h_{+}$, but
we allow for $v_{z}\sim 1$. We will also consider slow variations such that $%
\partial _{t}\ll \omega _{c}\equiv qB_{0}/m$ for each plasma species. With
these restrictions the Maxwell and fluid equations (\ref{me1})-(\ref{momeq1}%
) can be reduced to the following set of equations \cite{fotnot4}: 
\begin{eqnarray}
\partial _{z}E_{z} &=&\sum_{s}q\gamma (n_{0}+\delta n)  \label{me2} \\
\partial _{t}E_{y}-\partial _{z}B_{x} &=&-\sum_{s}\left( q\gamma
(n_{0}+\delta n)v_{y}\right) -{\textstyle\frac{1}{2}}E_{y}\partial _{t}h_{+}+%
{\textstyle\frac{1}{2}}(B_{0}+B_{x})\partial _{z}h_{+}  \label{ampeq} \\
\partial _{t}E_{z} &=&-\sum_{s}q\gamma \left( n_{0}+\delta n\right) v_{z} \\
\partial _{t}B_{x}-\partial _{z}E_{y} &=&-{\textstyle\frac{1}{2}}%
E_{y}\partial _{z}h_{+}+{\textstyle\frac{1}{2}}(B_{0}+B_{x})\partial
_{t}h_{+}  \label{faraday} \\
\partial _{t}\left( \gamma (n_{0}+\delta n)\right) &=& -\partial _{z}\left(\gamma (n_{0}+\delta n)v_{z}\right) \\
\partial _{t}(\gamma v_{y})+v_{z}\partial _{z}(\gamma v_{y}) &=&{\textstyle%
\frac{q}{m}}(E_{y}+v_{z}(B_{0}+B_{x})) \\
\partial _{t}(\gamma v_{z})+v_{z}\partial _{z}(\gamma v_{z}) &=&{\textstyle%
\frac{q}{m}}(E_{z}-v_{y}(B_{0}+B_{x}))  \label{fleq3}
\end{eqnarray}
and the linearized Einstein equations, (\ref{ee1122})-(\ref{ee1221}), become 
\begin{equation}
(\partial _{t}^{2}-\partial _{z}^{2})h_{+}=\kappa
(E_{y}^{2}-2B_{0}B_{x}-B_{x}^{2})  \label{gwevol}
\end{equation}
For the electromagnetic wave to exchange energy effectively with the
gravitational wave, the two waves must be almost resonant, so we will
consider all perturbations to be of the form $B_{x}\approx B_{x}(z-t)$,
which allows us to use $\partial _{t}\approx -\partial _{z}$ in simplifying
our system of equations. A certain care must be taken here though; this
approximation may of course not be used directly in the operator $(\partial
_{t}^{2}-\partial _{z}^{2})$. Equations (\ref{me2})-(\ref{fleq3}) can then
be reduced to 
\begin{eqnarray}
(\partial _{t}^{2}-\partial _{z}^{2})B_{x}+\sum_{s}\frac{m^{2}\omega _{p}^{2}%
}{q^{2}}\partial _{z}\left( \frac{\partial _{z}(\gamma v_{z})}{B_{0}+B_{x}}%
\right) &=&\partial _{t}((B_{x}+E_{y})\partial _{t}h_{+})+B_{0}\partial
_{t}^{2}h_{+}  \label{faramp} \\
\partial _{t}(\gamma v_{y})+v_{z}\partial _{z}(\gamma v_{y}) &=&{\textstyle%
\frac{q}{m}}(v_{z}(B_{0}+B_{x})+E_{y})  \label{v3eq}
\end{eqnarray}
where $\omega _{p}=\sqrt{n_{0}q^{2}/m}$ is the plasma frequency. In order to
eliminate $E_{y}$ we can start with the linearized version of Eq. (\ref
{faraday}). Using $\partial _{t}\approx -\partial _{z}$ and integrating we
obtain 
\begin{equation}
E_{y}=-B_{x}+{\textstyle\frac{1}{2}}B_{0}h_{+}  \label{ey}
\end{equation}
Re-inserting this into (\ref{faraday}) we find that the equation is
fulfilled to $\mathcal{O}(h_{+})$, which means that we can use the
expression (\ref{ey}) for $E_{y}$ even when the electromagnetic amplitude is
large, i.e. comparable to $B_{0}$.

We now make an expansion of Eq. (\ref{v3eq}) in the small parameter $%
\partial _{t}/\omega _{c}$ and use (\ref{ey}) to obtain $v_{z}$. The result
is 
\begin{equation}
v_{z}=\frac{B_{x}-{\textstyle\frac{1}{2}}B_{0}h_{+}}{B_{0}+B_{x}}+\mathcal{O}%
(\frac{\partial _{t}}{\omega _{c}})\approx \frac{B_{x}}{B_{0}+B_{x}}
\label{vz}
\end{equation}
We see that $v_{z}$ does not depend on the charge of the particles, but is
the same for all particle species, which means that the z-component of the
current is zero, and the current is in the y-direction. Inserting $v_{z}$, $%
E_{y}$ and expanding the wave operator in Eq. (\ref{faramp}) as $\partial
_{t}^{2}-\partial _{z}^{2}=(\partial _{t}-\partial _{z})(\partial
_{t}+\partial _{z})\approx 2(\partial _{t}+\partial _{z})\partial _{t}$ we
are left with the following coupled system of equations: 
\begin{eqnarray}
(\partial _{t}+\mathcal{V}(B_{x})\partial _{z})B_{x} &=&\frac{1}{2}%
B_{0}\partial _{t}h_{+}  \label{waveq} \\
(\partial _{t}^{2}-\partial _{z}^{2})h_{+} &=&-2\kappa B_{0}B_{x}
\label{eefinal}
\end{eqnarray}
where $\mathcal{V}(B_{x})=1-(1/2C_{A}^{2})\left( B_{0}/\left(
B_{0}+2B_{x}\right) \right) ^{3/2}$ and we have introduced the
Alfv\'{e}n velocity $C_{A}\equiv ( 1/\sum_{s}\omega
_{p}^{2}/\omega _{c}^{2})^{1/2}$. We note from (\ref{waveq})
that our previous assumption $\partial _{t}\approx -\partial _{z}$ requires
the background parameters to satisfy $C_{A}^{2}\gg 1$. In arriving at Eq. (%
\ref{eefinal}) we have also neglected a small term $-\kappa h_{+}B_{0}B_{x}$,
resulting from the substitution of (\ref{ey}) into (\ref{gwevol}). The left hand side of Eq. (\ref
{waveq}) contains the wave operator for compressional Alfv\'{e}n waves (also
called fast magnetosonic waves), which in the considered regime propagates
close to the speed of light. Furthermore, the left hand side of Eq. (\ref
{eefinal}) describes gravitational waves in vacuum. The right hand sides of
Eqs. (\ref{waveq}) and (\ref{eefinal}) are the mutual interaction terms,
which may provide a comparatively effective energy exchange, since the
propagation velocities of the wave modes are close to each other. Now
combining these two equations, and again expanding $\partial
_{t}^{2}-\partial _{z}^{2}$ we obtain, after one time integration, the
following wave equation for the combined electromagnetic and gravitational
wave mode: 
\begin{equation}
(\partial _{t}+\partial _{z})(\partial _{t}+\mathcal{V}(B_{x})\partial
_{z})B_{x}=-\frac{\kappa B_{0}^{2}}{2}B_{x}  \label{waveqfin}
\end{equation}

\subsection{WKB-approximation for Quasimonochromatic waves}\label{nlssect}

Here we intend to show that for quasimonochromatic waves, the wave equation (%
\ref{waveqfin}) leads to the nonlinear Schr\"{o}dinger (NLS) equation \cite
{doddeilbeck} for the weakly varying amplitude. We begin the analysis of
equation (\ref{waveqfin}) by first considering the linear case. Thus
replacing $\mathcal{V}(B_{x})$ with $\mathcal{V}(B_{0})=1-1/2C_{A}^{2}$ we
get 
\begin{equation}
(\partial _{t}+\partial _{z})\left( \partial _{t}+\left( 1-\frac{1}{%
2C_{A}^{2}}\right) \partial _{z}\right) B_{x}=-\frac{\kappa B_{0}^{2}}{2}%
B_{x}
\end{equation}
From a plane wave ansatz, $B_{x}=\bar{B}\exp (i(kz-\omega t))$%
, we then directly obtain the linear dispersion relation 
\begin{equation}
D(\omega ,k)=(\omega -k)(\omega -k+\frac{k}{2C_{A}^{2}})-\frac{\kappa
B_{0}^{2}}{2}=0  \label{disprel}
\end{equation}
This is in agreement with the dispersion relation presented by Ref. \cite
{bms}, if we in their result make the appropriate approximations
corresponding to $\partial _{t}\approx -\partial _{z}$. Solving the
dispersion relation we get the following result: 
\begin{equation}
\omega =k\left( 1-\frac{1}{4C_{A}^{2}}\right) \pm \sqrt{\frac{k^{2}}{%
16C_{A}^{4}}+\frac{\kappa B_{0}^{2}}{2}}  \label{disprelsol}
\end{equation}
We see that there are two roots of (\ref{disprel}) and we will hereafter
refer to these two modes as the fast mode, for the root with positive sign,
and the slow mode, for the root with negative sign. We see that for large $k$%
, i.e. $k\gg \sqrt{\kappa B_{0}^{2}}$ $C_{A}^{2}$, we can neglect the small
quantity $\kappa B_{0}^{2}/2$, and the dispersion relation for the fast mode
takes the form $\omega \approx k$, while for the slow mode we get $\omega
\approx k(1-1/2C_{A}^{2})$. We note that in this regime, the main part of
the energy is in the form of gravitational wave energy for the fast mode,
whereas most of the energy is electromagnetic for the slow mode. This can be
seen from the approximate dispersion relations above, together with the
coupled equations (\ref{waveq}) and (\ref{eefinal}). For very long
wavelengths, $k\lesssim $ $\sqrt{\kappa B_{0}^{2}}$ $C_{A}^{2}$, we of
course still have two roots of the dispersion relation, but in this case the
two modes divide their energies roughly equal between the gravitational and
electromagnetic form. Therefore we refer to this regime as that of mixed
modes.

The next step is to include higher order terms in the amplitude expansion of 
$\mathcal{V}(B_{x})$ and let the wave amplitude vary weakly in space and
time, as compared to $\omega $ and $k$. Keeping terms up to third order in $%
B_{x}$, Eq. (\ref{waveqfin}) now reads 
\begin{equation}
(\partial _{t}+\partial _{z})\left( \partial _{t}+\left( 1-\frac{1}{%
2C_{A}^{2}}+\frac{3}{2C_{A}^{2}B_{0}}B_{x}-\frac{15}{4C_{A}^{2}B_{0}^{2}}%
B_{x}^{2}\right) \partial _{z}\right) B_{x}=-\frac{\kappa B_{0}^{2}}{2}B_{x}
\label{3ordwaveq}
\end{equation}
We note that the nonlinear terms induces second harmonic (SH) and low
frequency (LF) perturbations, and we must therefore modify our ansatz
according to: 
\begin{equation}
B_{x}=B(z,t)e^{i(kz-\omega t)}+B_{SH}(z,t)e^{2i(kz-\omega
t)}+c.c.+B_{LF}(z,t)  \label{ansatz}
\end{equation}
where c.c. stands for complex conjugate of the preceding terms, and we
define $\omega $ and $k$ in (\ref{ansatz}) to fulfill the dispersion
relation (\ref{disprel}) exactly. Inserting this into (\ref{3ordwaveq}) we
get a long equation involving both SH- and LF-terms as well as terms of the
original frequency $\omega $. To analyze this we use standard techniques for
nonlinear wave equations, which we give an outline of here (see e.g. Ref. 
\cite{doddeilbeck} for details):

\begin{itemize}
\item  First we note that the induced SH- and LF-terms are one order smaller
in an amplitude expansion than terms of the original frequency, which allows
us to neglect all terms of higher order in $B_{SH}$ and $B_{LF}$.

\item  From linear theory we know that the envelope of the original
perturbation $B(z,t)$ travels with the group velocity $v_{g}$. Since
quadratic perturbations of the original amplitude acts as a driver for both
second harmonic and low-frequency perturbations, we observe that also $B_{SH}
$ as well as $B_{LF}$ will propagate with the group velocity. Thus we may
use the approximation $\partial _{t}\approx -v_{g}\partial _{z}$ for
derivatives acting on the \textit{amplitudes}. Note also that in general we
must use this approximation here instead of $\partial _{t}\approx -\partial
_{z}$. This is because of the operator $(\partial _{t}+\partial _{z})$,
which otherwise will give zero to lowest order when acting on the wave
perturbation.

\item  The different time scales in the equation (second harmonics, low
frequency, original frequency) are picked out by multiplying with the
appropriate exponential function and then averaging over many wavelengths and period times in
order to omit the rapidly oscillating terms in the resultant equation, see 
\cite{doddeilbeck}. (E.g. to find the equation governing the wave of
original frequency, $\omega $, we multiply by $\exp (-i(kz-\omega t))$ and
take an average over several wavelengths and period times.)
\end{itemize}

To lowest order, the equation for the SH-terms becomes 
\begin{equation}
D(2\omega ,2k)B_{SH}=\frac{3(\omega k-k^{2})}{B_{0}C_{A}^{2}}B^{2}
\label{sheq1}
\end{equation}
where, from (\ref{disprel}), we note that $D(2\omega ,2k)=3\kappa
B_{0}^{2}/2 $. Similarly, the equation for the low frequency wave is 
\begin{equation}
\left( \partial _{z}^{2}-C_{A}^{4}\left( 4\omega -4k+\frac{k}{C_{a}^{2}}%
\right) ^{2}\right) B_{LF}=\frac{3(4\omega -4k+k/C_{A}^{2})}{2B_{0}(\omega
-k+k/2C_{A}^{2})}\partial _{z}^{2}\lvert B\rvert ^{2}  \label{lfeq}
\end{equation}
where, in arriving at (\ref{lfeq}), we have used $\partial _{t}\approx
-v_{g}\partial _{z}$ and $v_{g}$ is given by 
\begin{equation}
v_{g}=\frac{2\omega (1-1/4C_{A}^{2})-2k(1-1/2C_{A}^{2})}{2\omega
-2k(1-1/4C_{A}^{2})}
\end{equation}
Note that if $C_{A}^{2}\rightarrow \infty $, i.e. the plasma density goes to
zero, $v_{g}\rightarrow 1$, although, as can be seen from (\ref{disprel}), $%
\omega -k\neq 0$.

The back reaction at the original frequency is determined by the equation 
\begin{equation}
\left( i\left( \partial _{t}+v_{g}\partial _{z}\right) +\frac{v_{g}^{\prime }%
}{2}\partial _{z}^{2}\right) B=\frac{F}{B_{0}}\left( B_{SH}B^{\ast }+B_{LF}B+%
\frac{5}{2B_{0}}\lvert B\rvert ^{2}B\right)  \label{orgeq}
\end{equation}
where the group dispersion, $v_{g}^{\prime }$, is given by 
\begin{eqnarray}
\frac{v_{g}^{\prime }}{2} &=&\frac{\kappa B_{0}^{2}}{C_{A}^{4}\left( 4\omega
-4k+{\textstyle\frac{k}{C_{A}^{2}}}\right) ^{3}}, \\
F &=&\frac{3(k^{2}-\omega k)}{2C_{A}^{2}\left( 2k-2\omega -\frac{k}{%
2C_{A}^{2}}\right) }
\end{eqnarray}
and the star denotes complex conjugate.

In order to simplify the system of equations (\ref{sheq1})-(\ref{orgeq}), we
first consider the LF-equation (\ref{lfeq}). We see that for the different
modes we can compare the solution of equation (\ref{lfeq}) with the SH-term
and get $B_{LF}/B_{SH}\sim \partial _{z}^{2}|B|^{2}/k^{2}$. Now remembering
that the wave amplitudes vary weakly in space, as compared to $k$, we then
conclude that this quantity is very small and we therefore neglect the term
involving the low frequency field from now on.

Let us now write equation (\ref{orgeq}) in a more simple form by using (\ref
{sheq1}) to eliminate $B_{SH}$ and making the following coordinate
transformations 
\begin{eqnarray}
\xi &=&\beta (z-v_{g}t)  \label{xitrans} \\
\tau &=&\frac{\left\vert v_{g}^{\prime }\right\vert \beta ^{2}}{2}t
\label{tautrans}
\end{eqnarray}
where 
\begin{equation}
\beta =C_{A}^{2}\left( 4\omega -4k+{\textstyle\frac{k}{C_{A}^{2}}}\right)
\end{equation}
and the absolute value in the transformation (\ref{tautrans}) has been
introduced in order to keep $\tau$ positive at all positive times, $t$. Note
that the sign of $v_{g}^{\prime}$ depends on the roots of the dispersion
relation; for the mode with positive sign in (\ref{disprelsol}) $%
v_{g}^{\prime}$ is positive and for the mode with negative sign $%
v_{g}^{\prime}$ is negative. Furthermore we use the linearized equations to
relate $B$ to the gravitational perturbation $h_{+}$. Keeping only lowest
order terms in Eq. (\ref{waveq}) we get 
\begin{equation}
h_{+}=\frac{2\left( \omega -k+{\textstyle\frac{k}{2C_{A}^{2}}}\right) }{%
B_{0}\omega }B  \label{btoh}
\end{equation}
Transforming the gravitational perturbation amplitude according to 
\begin{equation}
\tilde{h}_{+}=\Psi _{1}h_{+}  \label{htrans}
\end{equation}
where 
\begin{equation}
\Psi _{1}^{2}=\frac{3\omega ^{2}k\left( 5\omega -5k+{\textstyle\frac{9k}{%
2C_{A}^{2}}}\right) }{16C_{A}^{2}\left( \omega -k+{\textstyle\frac{k}{%
2C_{A}^{2}}}\right) ^{4}}
\end{equation}
then allows us to rewrite Eq. (\ref{orgeq}) as a standard NLS equation for
the rescaled gravitational perturbation amplitude: 
\begin{equation}
\left( i\partial _{\tau }\pm\partial _{\xi }^{2}\right) \tilde{h}%
_{+}=\pm\lvert \tilde{h}_{+}\rvert ^{2}\tilde{h}  \label{NLS-norm}
\end{equation}
Here the plus and minus sign correspond to the fast and the slow mode
respectively. Eq. (\ref{NLS-norm}) can be solved by the inverse scattering
technique, as discussed by for example Ref. \cite{doddeilbeck}. Asymptotically in time, the solution is a train of solitons.

\subsection{Generalization to a 3D spatial dependence}\label{3dsect}

In this section we address the fact that in reality we will not
have exact plane wave solutions to the linearized Einstein and Maxwell
equations. Rather the involved quantities will also depend on the x- and
y-coordinates if we have wave propagation in the z-direction. We will here
assume that this deviation from plane waves is small, and use a perturbative
treatment. The underlying assumption is that $\partial _{x}$ and $\partial
_{y}$ is of order of $1/L_{\perp}$, where $1/L_{\perp}$ is the characteristic width of the pulse fulfilling $\lambda /L_{\perp}\ll 1$ ($\lambda $ is the wavelength), while $\partial _{z}$ is of order $1/\lambda $.

Assuming an x- and y-dependence in the metric perturbation function, $h_{+}$%
, will induce perturbations in the form of additional components in the
Einstein tensor, as seen from the Lorentz gauge condition (\ref{lorentz}). We thus note that there must be corresponding perturbations in
the energy-momentum tensor in order to fulfill Einstein's equations (\ref
{pertefe}). The metric must of course be adjusted correspondingly, and this
leads to the addition of components in the metric perturbation, $h_{ab}$;
specifically the following components (of order $\lambda /L_{\perp}$) must be added: $h_{01}$, $h_{02}$, $%
h_{13}$ and $h_{23}$. Inserting this metric perturbation into (\ref{pertefe}%
) and using the Lorentz-condition 
\begin{equation}\label{lorentz}
\partial_{b}h^{ab}=0
\end{equation}
we get the following modified version of Eq. (\ref{ee1122}): 
\begin{equation}  \label{ee1122perp}
(\partial_{t}^{2}-\partial_{z}^{2}-\nabla_{\perp}^{2})h_{+}=\kappa(\delta
T_{11}-\delta T_{22})
\end{equation}
where we again have subtracted the $_{11}$ and $_{22}$ components of Eq. (%
\ref{pertefe}), and we have introduced $\nabla_{\perp}^{2}\equiv%
\partial_{x}^{2}+\partial_{y}^{2}$.

The variation of the
perturbed fields in the x- and y-directions will also induce other
components of the electromagnetic and velocity field perturbations, so we
introduce additional terms in the perturbed fields according to $\mathbf{E}%
=E_{x}\mathbf{e}_{1}+E_{y}\mathbf{e}_{2}+E_{z}\mathbf{e}_{3}$ and
correspondingly for other fields. These extra perturbations are small, of order $\lambda/L_{\perp}$. Note that we now have several small quantities here and we neglect
terms of the type $\partial _{x}h_{+}$, $\partial _{x}E_{x}$, $v_{x}B_{y}$
etc., i.e. we neglect all terms that include more than one small quantity
(with the important exception of the $\nabla _{\perp }^{2}$-operator in (\ref
{ee1122perp}), because this operator compares with the
operator $(\partial _{t}^{2}-\partial _{z}^{2})$, which is small in
itself). We also make the same approximations made previously.

With these approximations we can calculate the effective charges, currents
and forces (\ref{roe1})-(\ref{gforce1}) and work out the Maxwell and fluid
equations (\ref{me1})-(\ref{momeq1}). The result is a large set of
equations, that can be reduced using $\partial_{z}\approx-\partial_{t}$, and
the first order results (\ref{ey}) and (\ref{vz}). We finally arrive at the
following equation for the magnetic field perturbation
\begin{equation}  \label{beqperp}
\left( \partial_{t}^{2}-\partial_{z}^{2}-\nabla_{\perp}^{2}\right) B_{x}+%
\frac{1}{C_{A}^{2}}\partial_{z}\left(\left(\frac{B_{0}}{B_{0}+2B_{x}}%
\right)^{\frac{3}{2}}\partial_{z}B_{x}\right)=B_{0}\partial_{t}^{2}h_{+}
\end{equation}
and Eq. (\ref{ee1122perp}) reduces to 
\begin{equation}  \label{ee1122perp2}
(\partial_{t}^{2}-\partial_{z}^{2}-\nabla_{\perp}^{2})h_{+}=-2\kappa
B_{0}B_{x}
\end{equation}
In order to simplify this system further we expand the operator $%
(\partial_{t}^{2}-\partial_{z}^{2}-\nabla_{\perp}^{2})\approx2\partial_{t}(%
\partial_{t}+\partial_{z})-\nabla_{\perp}^{2}$ and insert this in (\ref
{beqperp})-(\ref{ee1122perp2}). Let us also introduce the notation $%
\partial_{t}^{-1}$ for the inverse operation of $\partial_{t}$. Using (\ref
{ee1122perp2}) we can then combine these two equations into the following
modified version of the wave equation (\ref{waveqfin}) 
\begin{equation}  \label{beqperp2}
\left(\partial_{t}+\partial_{z}-\frac{1}{2}\partial_{t}^{-1}\nabla_{%
\perp}^{2}\right)\left(\partial_{t}+\mathcal{V}(B_{x})\partial_{z}-\frac{1}{2%
}\partial_{t}^{-1}\nabla_{\perp}^{2}\right) B_{x}=-\frac{\kappa B_{0}^{2}}{2}%
B_{x}
\end{equation}

Next we note that for perturbations on the form $B_{x}\sim B(t)\exp {%
i(kz-\omega t)}$ we can expand the operator $\partial _{t}$ as $\partial
_{t}=-i\omega +\tilde{\partial}_{t}$, where $\tilde{\partial}_{t}$ acts only
on the amplitude. If $B(t)$ varies slowly compared to the exponential part
(i.e. $\frac{\tilde{\partial}_{t}}{\omega }\ll 1$), we can then expand the
inverse operator, $\partial _{t}^{-1}$ as 
\begin{equation}
\partial _{t}^{-1}=\frac{1}{-i\omega +\tilde{\partial}_{t}}=\frac{i}{\omega }%
\left( 1-i\frac{\tilde{\partial}_{t}}{\omega }+\ldots \right) \approx \frac{i%
}{\omega }
\end{equation}
Using this result and expanding the velocity function $\mathcal{V}(B_{x})$
up to second order in $B_{x}$ we get the equation 
\begin{equation}
\left( \partial _{t}+\partial _{z}\right) \left( \partial _{t}+\mathcal{V}%
_{exp}(B_{x})\partial _{z}\right) B_{x}-\left( \frac{i}{\omega }\left(
\partial _{t}+\partial _{z}-\frac{1}{4C_{A}^{2}}\partial _{z}\right) \nabla
_{\perp }^{2}\right) B_{x}=-\frac{\kappa B_{0}^{2}}{2}B_{x}
\label{3ordwaveqperp}
\end{equation}
where $\mathcal{V}_{exp}(B_{x})=1-(1/2C_{A}^{2})+(3B_{x}/2C_{A}^{2}B_{0})-(15B_{x}^{2}/4C_{A}^{2}B_{0}^{2})$. We see that
this is just equation (\ref{3ordwaveq}) with an extra term on the left hand
side corresponding to the small variation in the x- and y-directions. In
order to analyze this equation we make a small modification to our ansatz (%
\ref{ansatz}) in order to include the weak spatial dependence in the
amplitudes and insert this into (\ref{3ordwaveqperp}). Working out the
algebra, we see that the equations for the SH- and LF-terms are unaltered to
the desired accuracy. This means that the only alteration to our previous
results in section \ref{nlssect} will be an extra term in the operator on
the left hand side of equation (\ref{orgeq}), corresponding to the $%
\nabla_{\perp}^{2}$-term in (\ref{3ordwaveqperp}). Making the same
transformations made in section \ref{nlssect}, the final result for the
transformed gravitational perturbation amplitude, $\tilde{h}_{+}$, is: 
\begin{equation}
\left( i\partial _{\tau }\pm\partial _{\xi }^{2}+\Upsilon \nabla _{\perp
}^{2}\right) \tilde{h}_{+}=\pm\lvert \tilde{h}_{+}\rvert ^{2}\tilde{h}_{+}
\label{NLS-3D}
\end{equation}
where the coefficient $\Upsilon $ is given by 
\begin{equation}
\Upsilon =\frac{\left\vert\left( 4\omega -4k+{\textstyle\frac{k}{C_{A}^{2}}}%
\right)\right\vert}{4\omega (\omega -k)\left( \omega -k+{\textstyle\frac{k}{%
2C_{A}^{2}}}\right) }
\end{equation}
and where the plus and minus signs again refer to the fast and slow mode
respectively.

\section{Summary and discussion}

We have studied the weakly nonlinear propagation of coupled Alfv\'{e}n and
gravitational waves (AGW) propagating perpendicular to an external magnetic
field in a plasma with $C_{A}^{2}\gg 1$, using the coupled Einstein and
Maxwell equations. One of our main results is the general evolution equation
(\ref{waveqfin}), which holds for a broad band spectrum. A thorough study of
this equation is a project for further study. In order to simplify (\ref
{waveqfin}), we have made a WKB-ansatz, and showed that this leads to the
well-known NLS-equation (\ref{NLS-norm}), \ where the generalization to a 3D
spatial dependence has been described in section \ref{3dsect}.

Possible astrophysical sources for large amplitude AGW are binary pulsars,
particularly in the later stages of their evolution, i.e. not too far from
merging. Clearly such a system produces gravitational waves, and, as
discussed by for example Ref. \cite{bmd1}, this leads also to large
amplitude electromagnetic perturbations. Through the back reaction in EE the
electromagnetic perturbation is coupled to the gravitational one, and
linearly this is described by the dispersion relation (\ref{disprel}). For
typical wavelengths and background parameters (i.e. for $k\gg \sqrt{\kappa
B_{0}^{2}}$ $C_{A}^{2}$), the dispersion relation approximately separates in
the fast mode, $\omega \approx k$, and the slow mode, $\omega \approx
k(1-1/2C_{A}^{2})$, where the former mainly has gravitational character and
the later one electromagnetic character. However, very close to collapsing
binaries we have magnetic fields $B_{0}\sim 10^{8}\mathrm{T}$ , in which
case there is no clear distinction in character between the modes. On the
other hand, for moderate values of $B_{0}$ further from the source, the
modes will be clearly separated. However, due to the strong coupling in the
near region of pulsars, both the fast and slow mode will be represented at
larger distances, although the energy density of the later will be smaller.
The amount of energy given to the slow mode can be estimated from the
results of Ref. \cite{bmd1}, and correspond to an electromagnetic power of
the order $10^{25}$ $\mathrm{W}$ close to merging.

In order to discuss the nonlinear evolution, we consider pulses including a
perpendicular dependence that is described by Eq. (\ref{NLS-3D}). For simplicity
we study long pulses with a shape that depend only on a (normalized)
cylindrical radius, $\tilde{r}=\sqrt{(x^{2}+y^{2})/\Upsilon }$, thus
neglecting dispersive effects. On reinstating the speed of light $c$, the
condition for this is $16\pi GB_{0}^{2}C_{A}^{2}L_{\perp}^{2}/c^{6}\mu
_{0}\ll k^{2}L^{2}_{\parallel}$, where $L_{\perp}$ and $L_{\parallel}$ are the characteristic length scales for
variations in the perpendicular and parallel directions respectively. In this case we get a
cylindrically symmetric NLS equation 
\begin{equation}
\left( i\partial _{\tau }+\frac{1}{\tilde{r}}\frac{\partial }{\partial 
\tilde{r}}\left( \tilde{r}\frac{\partial }{\partial \tilde{r}}\right)
\right) \tilde{h}_{+}=\pm \lvert \tilde{h}_{+}\rvert ^{2}\tilde{h}_{+}
\label{NLS-cyl}
\end{equation}
Contrary to Eq. (\ref{NLS-norm}), there are no exact solutions known for the
cylindrical version (\ref{NLS-cyl}) , except for some physically
uninteresting special cases. Still Eq. (\ref{NLS-cyl}) has been studied in
some detail, both analytically, using approximate variational techniques 
\cite{Desaix91}, and numerically \cite{Kuznetsov86}. Of most interest to us
is the case with the minus-sign, corresponding to the slow,
electromagnetically dominated mode. The main motivation for considering this
mode, is that this choice gives a nonlinearity of focusing type. For strong
enough nonlinearity the solutions to (\ref{NLS-cyl}) then show wave
collapse, i.e. the pulse focuses indefinitely and the pulse radius $\tilde{r}%
_{p}\rightarrow 0$ in a finite time. The main characteristics of the
collapse can be described within an approximate variational framework.
Following Ref. \cite{Desaix91} we use a trial function $\tilde{h}_{+}(\tilde{r}%
,\tau )=A(\tau )\mathrm{sech}(\tilde{r}/a(\tau ))\exp (\mathrm{i}b(\tau )%
\tilde{r}^{2})$ together with Rayleigh-Ritz optimization in order to derive
a single ordinary differential equation for the pulse width $a(\tau )$. \
Pulse energy conservation then follows from the optimization scheme
according to $a(\tau )A(\tau )=A(0)a(0)$. Furthermore, the evolution of the
width scales as $[a(\tau )(a(0)]^{2}-1\propto (1-A(0)^{2}a(0)^{2}/I_{c})\tau
^{2}$ where $I_{c}\approx 1.35$. Thus if $A(0)^{2}a(0)^{2}>1.35$, the pulse
will collapse to zero width in a finite time, and in the opposite case
linear diffraction will dominate to spread out the pulse indefinitely. Of
course, for the case of collapse, higher order nonlinearities neglected in
the derivation of (\ref{NLS-cyl}) will eventually become important, which
will change the later stages of the given scenario.

Let us study whether the electromagnetically dominated slow mode, excited by
binary pulsars close to collapse may undergo wave collapse. Here a word of caution is at hand. In a real system of this type 
a number of effects outside the model equation (\ref{NLS-cyl}) are likely
to play important roles for the pulse dynamics. For the sake of
simplicity we will here exclude such effects. From the variational approach mentioned above, which agrees with numerical works \cite{Kuznetsov86}, we find that collapse takes place if $\lvert \tilde{h}%
_{+}\rvert ^{2}>I_{c}/\tilde{r}_{p}^{2}$, which can be written 
\begin{equation}
\frac{B^{2}}{B_{0}^{2}}\frac{c^{6}}{C_{A}^{6}}\frac{3\mu _{0}k^{2}c^{4}}{%
8\pi GB_{0}^{2}}k^{2}r_{p}^{2}>1.35  \label{Collapsen-ny-cond}
\end{equation}
where we have reintroduced dimensional quantities. The electromagnetic
fields leaving the binary system is excited by gravitational quadrupole
radiation, which has a certain directionality, but not a very pronounced
one. Thus before significant focusing takes place, the pulse width can be
replaced by the radial distance from the source, $r_{\mathrm{dist}}$, as a
rough order of magnitude estimate for the pulse width. \ Thus we substitute $%
r_{p}\rightarrow r_{\mathrm{dist}}$ in (\ref{Collapsen-ny-cond}), use $%
B^{2}r_{\mathrm{dist}}^{2}=10^{25}\mu _{0}/c4\pi $ $\mathrm{Tm}^{2}$ and
take $k=10^{-5}$ $\mathrm{m}^{-1}$ corresponding to the parameter values
mentioned above, discussed in more detail in Ref. \cite{bmd1}. The only
extra parameters we need to specify are $B_{0}$ and $\rho _{0}$. For a broad
range of relevant background values $10^{-20}\ \mathrm{kg/m}^{3}<\rho
_{0}<10^{-5}\ \mathrm{kg/m}^{3}$ and $10^{-10}\ \mathrm{T}<B_{0}<0.1\ 
\mathrm{T}$\textrm{,} the condition (\ref{Collapsen-ny-cond}) is always
fulfilled. The main concern is to find values fulfilling the previous
assumption $C_{A}^{2}/c^{2}=B_{0}^{2}/\mu _{0}\rho _{0}c^{2}\gg 1$, which
holds for a low-density plasma $\rho _{0}\sim 10^{-20}\ \mathrm{kg/m}^{3}$,
provided $B_{0}>10^{-4}\ \mathrm{T}.$ However, we note that the condition (%
\ref{Collapsen-ny-cond}) can be misleading, since for the strong powers
considered, the level of the second harmonic fields typically obeys $%
B_{SH}>B_{0}$ at relevant distances from the source. This level lies outside
the regime of validity for weakly nonlinear waves. We note that the left
hand side of (\ref{Collapsen-ny-cond}) is proportional to $B_{SH}$.
Replacing the expression for $B_{SH}$ found from (\ref{sheq1}) by its limit
of validity $B_{SH}\sim B_{0}$ (relevant for more moderate powers than $%
10^{25}\ \mathrm{W}$), we note that collapse will occur when 
\begin{equation}
\frac{c^{2}k^{2}r_{\mathrm{dist}}^{2}}{C_{A}^{2}}\gtrsim 1
\label{Collapse-critical}
\end{equation}
which is easily fulfilled at reasonable distances ($r_{\mathrm{dist}}\gtrsim
10^{7}\ \mathrm{m}$) from the source. The characteristic timescale for
significant focusing corresponding to the example given above (with $%
B_{SH}\sim B_{0}$) is given by $T_{\mathrm{foc}}\sim C_{A}^{2}/c^{2}\omega $.

The nonlinearity of the fast (gravitationally dominated) mode is not of
focusing type. We note that from Eq. (\ref{NLS-norm}) there is still the
possibility of other nonlinear effects, such as (dark) soliton formation
that might, in principle, be detected. The characteristic time scale for
such nonlinear processes is typically very large \cite{fotnot5}, which can be seen from eqs. (\ref{tautrans}%
)-(\ref{NLS-norm}) inserting realistic background parameter values for the
relevant quantities. This means that for nonlinear effects to be important
for this mode, we must assume extreme parameter values which cannot be
justified by astrophysical observations.

The condition (\ref{Collapse-critical}) does not contain gravitational
parameters, which is a consequence of the electromagnetic dominance of the
slow mode. However, we note that the considered process is induced by the
gravitational-electromagnetic coupling, and thus it still has a
gravitational origin. The condition for wave collapse of this mode can be
fulfilled for a reasonable range of parameters, which opens up for the
possibility of structure formation of the electromagnetic radiation pattern.
In the example considered above, the focusing takes place on fractions of a
second (unless $C_{A}^{2}$ is extremely large), and thus the nonlinearities
may cause noticeable structures in the later stages of binary merging. We
wish to point out here, though, that this example is to be seen as a
somewhat crude estimation of the focusing effect, and that this example do
not express all of the physics involved. In the absolute vicinity of a
binary merger, other significant effects are almost certain to appear. A
more complete treatment is a project for future research.

\acknowledgements{This work was supported by the Swedish National Graduate School of Space Technology.}


\begin{thebibliography}{99}
\bibitem{fotnot1} Here it is assumed that the different fluid species only interact through
the collective electromagnetic field, and otherwise the different fluids can
pass freely through each other. This is true in a plasma, where the
characteristic length scales for variations are much smaller than the
collisional mean free path.
\bibitem{fotnot2} Note here that the components $B_{x}$, $E_{y}$ etc. still refers to the
basis (\ref{tetrad}), and should not be confused with field components
referring to a coordinate basis. Although the difference between tetrad
components and coordinate components is very small, this difference may
still be significant in certain situations.
\bibitem{fotnot3} Actually there exists a weak coupling to $h_{\times }$, but working out the
equations we see that $h_{\times }\sim E_{x}E_{y}$ while $h_{+}\sim
B_{0}B_{x}$. The interaction between $h_{\times }$, $B_{y}$ and $E_{x}$ is
non-resonant so that $B_{y},E_{x}\sim h_{\times }$ will not grow, which
means that the coupling to $h_{+}$ is much stronger than to $h_{\times }$.
\bibitem{fotnot4} Note that, provided that the constraint equation (\ref{me2}) is initially
fulfilled, the equations (\ref{ampeq})-(\ref{fleq3}) make sure that (\ref
{me2}) is always fulfilled.
\bibitem{fotnot5} Note that there exist astrophysical background parameter values that may
give short nonlinear timescales, for example the environment close to
magnetars, but typically those regions occupy too small volumes to be of
real relevance in this context.
\end{thebibliography}
\end{document}